\documentclass[12pt]{article}
\usepackage{amsmath}
\usepackage{graphicx}

\newsavebox{\SKpath}
\sbox{\SKpath}{\includegraphics[width=8.8cm]{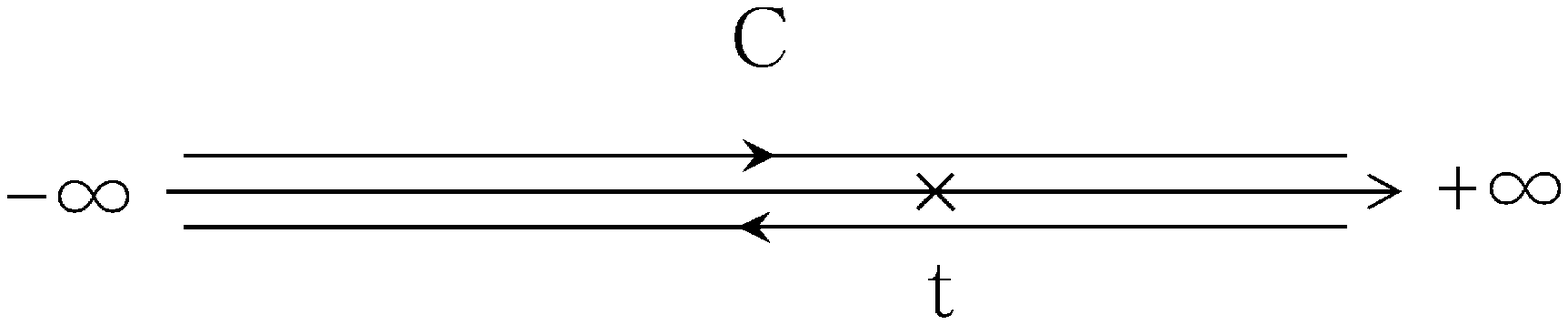}} 
\newlength{\SKpathl}
\newsavebox{\ThreeA}
\sbox{\ThreeA}{\includegraphics[width=4cm]{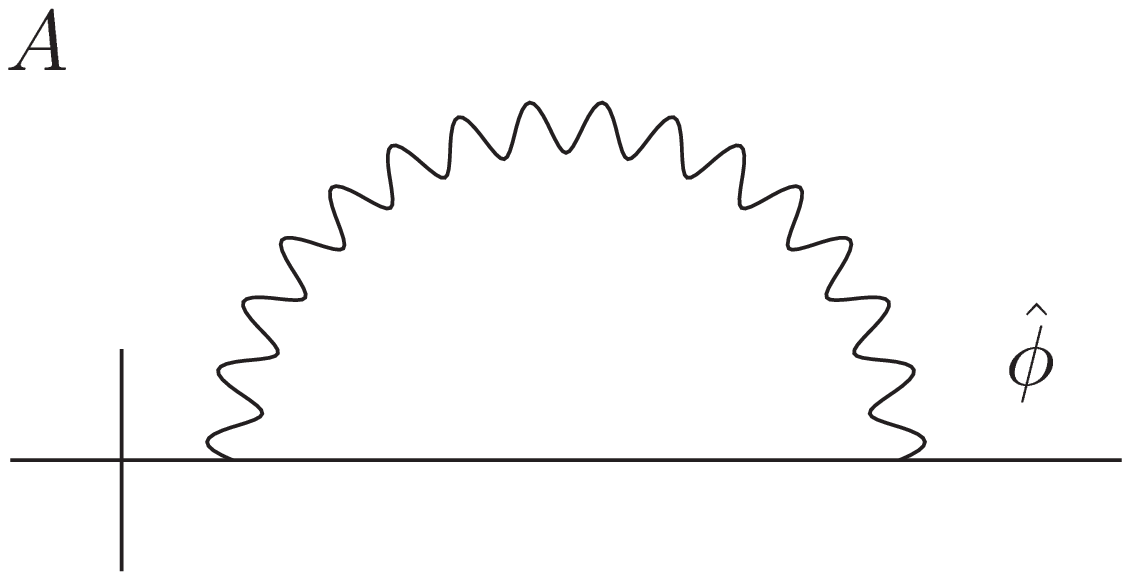}} 
\newlength{\ThreeAl}
\newsavebox{\ThreeB}
\sbox{\ThreeB}{\includegraphics[width=4cm]{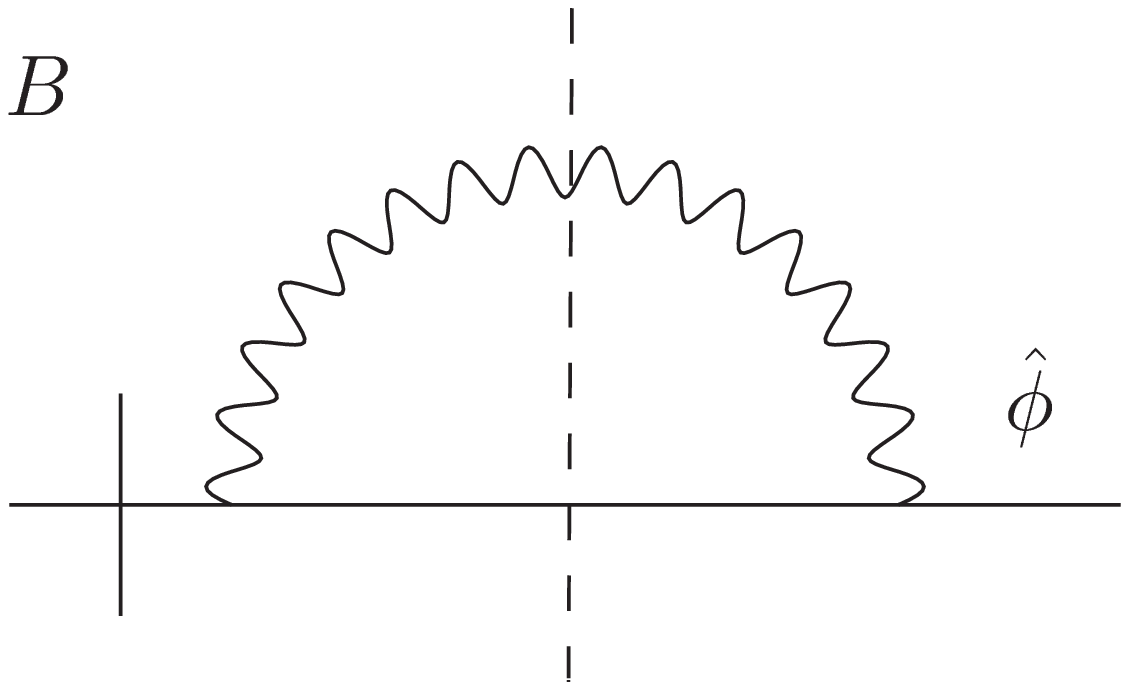}} 
\newlength{\ThreeBl}
\def\papertitlepage{\baselineskip 3.5ex\thispagestyle{empty}}
\def\preprinumber#1#2{\hfill\begin{minipage}{4.2cm} #1
        \par\noindent #2 \end{minipage}}
\allowdisplaybreaks[3]

\begin{document}

\papertitlepage
\setcounter{page}{0}
\preprinumber{KEK-TH-1591}{}
\baselineskip 0.8cm
\vspace*{2.0cm}

\begin{center}
{\Large\bf Scheme dependence of quantum gravity\\
on de Sitter background}
\end{center}

\begin{center}
Hiroyuki K{\sc itamoto}$^{1)}$\footnote{E-mail address: kitamoto@post.kek.jp} and
Yoshihisa K{\sc itazawa}$^{2),3)}$\footnote{E-mail address: kitazawa@post.kek.jp}\\
\vspace{5mm}
$^{1)}${\it Department of Physics and Astronomy\\
Seoul National University, 
Seoul 151-747, Korea}\\
$^{2)}${\it KEK Theory Center, 
Tsukuba, Ibaraki 305-0801, Japan}\\
$^{3)}${\it Department of Particle and Nuclear Physics\\
The Graduate University for Advanced Studies (Sokendai)\\ 
Tsukuba, Ibaraki 305-0801, Japan}
\end{center}

\vskip 5ex
\baselineskip = 2.5 ex

\begin{center}{\bf Abstract}\end{center}
We extend our investigation of the IR effects on the local dynamics of matter fields in quantum gravity. 
Specifically we clarify how the IR effects depend on the change of the quantization scheme: 
different parametrization of the metric and the matter field redefinition. 
Conformal invariance implies effective Lorentz invariance of the matter system in de Sitter space.
An arbitrary choice of the parametrization of the metric and the matter field redefinition does not preserve the effective Lorentz invariance of the local dynamics. 
As for the effect of different parametrization of the metric alone, the effective Lorentz symmetry breaking term can be eliminated by shifting the background metric. 
In contrast, we cannot compensate the matter field redefinition dependence by such a way. 
The effective Lorentz invariance can be retained only when we adopt the specific matter field redefinitions where all dimensionless couplings become scale invariant at the classical level. 
This scheme is also singled out by unitarity as the kinetic terms are canonically normalized.

\vspace*{\fill}
\noindent
May 2013

\newpage
\section{Introduction}
\setcounter{equation}{0}

In de Sitter (dS) space, the degrees of freedom at the super-horizon scale increase with cosmic expansion. 
This increase leads to the dS symmetry breaking term in the propagator of a massless and minimally coupled
scalar field and gravitational field. 
The symmetry breaking term is a direct consequence of the scale invariant spectrum and 
depends logarithmically on the scale factor of the universe \cite{Vilenkin1982,Linde1982,Starobinsky1982}. 
So in some field theoretic models in dS space, physical quantities become time dependent through the propagator. 

For a general scalar field theory, we need to fine-tune the mass term to obtain such infra-red (IR) effects. 
On the other hand, the gravitational field contains massless and minimally coupled modes without the fine-tuning. 
In this regard, the gravitational field is an attractive candidate which induces the IR effects. 
Since the product of the gravitational constant $\kappa^2$ and the Hubble constant $H$ is small in most cases of physical interest: $\kappa^2H^2\ll 1$, 
the quantum effects from gravity seem to be suppressed by the coupling. 
However, they are associated with the growing time dependence as $(\kappa^2H^2\log a(t))^n$, $a(t) = e^{Ht}$ at the $n$-loop level. 
It indicates that at late times, the IR effects may grow up to certain values which are not suppressed by the coupling. 

Observationally we live in a de Sitter like Universe while Lorentz invariance is well respected in the laboratory. 
For examples, let us consider a massless and conformally coupled scalar and massless Dirac field. 
At the classical level, Lorentz invariance follows from dS invariance due to the conformal symmetry. 
At the sub-horizon scale, the Lorentz invariance holds in non-conformally coupled field theories effectively. 
It is an important question to ask whether the IR effects from quantum gravity respect the effective Lorentz invariance of microscopic physics or not. 

We have investigated soft gravitational effects on the local dynamics of matter fields 
at the sub-horizon scale \cite{KitamotoSD,KitamotoG}. 
Although we cannot observe the super-horizon modes directly, 
it is possible that virtual gravitons of the super-horizon scale affect microscopic physics which are directly observable. 
These investigations have been performed mainly on the gauge introduced in \cite{Tsamis1992}. 
We have shown that the IR effects respect the effective Lorentz invariance in scalar, Dirac and gauge field theories at the one-loop level. 
It indicates that soft gravitational effects on free field theories can be absorbed by wave function renormalization factors. 
In the interacting field theories with $\phi^4$, Yukawa and gauge couplings, 
the dimensionless couplings are dynamically screened by soft gravitons. 
Even when we deform the gauge fixing term slightly, the effective Lorenz invariance of the local dynamics is preserved. 
Although the time dependence of each coupling is gauge dependent, their relative scaling exponents are gauge invariant. 

In addition to the gauge dependence, the quantum gravitational effects depend on the parametrization scheme of the metric. 
In \cite{Kahya2007,Miao2005(1),Miao2005(2),Miao2005(3)}, 
soft gravitational effects on free field theories have been investigated in the same gauge, 
but in a different parametrization of the metric from ours. 
There is also a difference from ours in the choice of the matter field redefinition. 
Actually the results obtained in these papers do not coincide with ours. 
In particular, the IR effect on a free Dirac field breaks the effective Lorentz invariance. 
We show the discrepancy originates just from the choice of the parametrization of the metric and the matter field redefinition. 

Since we do not observe the breakdown of the effective Lorentz invariance in our quantization scheme, 
there should be a prescription to retain it in a different quantization scheme. 
There must be a reasoning to select which quantization scheme should be chosen for the description of physics. 
In order to answer these questions, 
we clarify how the IR effects on the local dynamics depend on the parametrization of the metric and the matter field redefinition. 

The organization of this paper is as follows. 
In Section $2$, we quantize the gravitational field on the dS background. 
We identify the graviton modes which induces the dS symmetry breaking. 
In Section $3$, we clarify the parametrization dependence of soft gravitational effects on the dS background. 
Specifically, we derive the results obtained in \cite{Kahya2007,Miao2005(1),Miao2005(2),Miao2005(3)} 
by accounting the quantization scheme difference from ours \cite{KitamotoSD,KitamotoG}. 
The discussion in Section $4$ is the main topic of this paper. 
Here we report the prescription to retain the effective Lorentz invariance and 
the reasoning to select a quantization scheme which is appropriate for the description of physics. 
We conclude with discussions in Section $5$.

\section{Gravitational field in dS space}
\setcounter{equation}{0}

In this section, we review the gravitational field in dS space.  
In the Poincar\'{e} coordinate, the metric in dS space is
\begin{align}
ds^2=-dt^2+a^2(t)d{\bf x}^2,\hspace{1em}a(t)=e^{Ht}, 
\end{align}
where the dimension of dS space is taken as $D=4$ and $H$ is the Hubble constant. 
In the conformally flat coordinate,
\begin{align}
(g_{\mu\nu})_\text{dS}=a^2(\tau)\eta_{\mu\nu},\hspace{1em}a(\tau)=-\frac{1}{H\tau}. 
\end{align}
Here the conformal time $\tau\ (-\infty <\tau < 0)$ is related to the cosmic time $t$ as $\tau\equiv-\frac{1}{H}e^{-Ht}$. 
We assume that dS space begins at an initial time $t_i$ with a finite spacial extension.
After a sufficient exponential expansion, the dS space is well described locally by the above metric irrespective of the spacial topology. 

In our previous studies \cite{KitamotoSD,KitamotoG}, 
we have adopted the following parametrization of the metric: 
\begin{align}
g_{\mu\nu}=\Omega^2(x)\tilde{g}_{\mu\nu},\hspace{1em}\Omega(x)=a(\tau)e^{\kappa w(x)}, 
\label{para1}\end{align}
\begin{align}
\det \tilde{g}_{\mu\nu}=-1,\hspace{1em}\tilde{g}_{\mu\nu}=\eta_{\mu\rho}(e^{\kappa h(x)})^{\rho}_{~\nu}, 
\label{para2}\end{align}
where $\kappa$ is defined by the Newton's constant $G$ as $\kappa^2=16\pi G$. 
To satisfy (\ref{para2}), $h_{\mu\nu}$ is taken to be traceless
\begin{align}
h^{\mu}_{\ \mu}=0. 
\label{para3}\end{align}
On the other hand, 
R.P. Woodard and his collaborators have adopted a different parametrization 
in \cite{Tsamis1992,Kahya2007,Miao2005(1),Miao2005(2),Miao2005(3)}: 
\begin{align}
g_{\mu\nu}=a^2(\tau)(\eta_{\mu\nu}+2\kappa\Phi(x)\eta_{\mu\nu}+\kappa\Psi_{\mu\nu}(x)). 
\label{Wpara1}\end{align}
To facilitate the comparison with our parametrization (\ref{para1})-(\ref{para3}),
we have divided the fluctuation into the trace and traceless part
\begin{align}
\Psi^\mu_{\ \mu}=0. 
\label{Wpara2}\end{align}
That is, $w$, $h_{\mu\nu}$ are equal to $\Phi$, $\Psi_{\mu\nu}$ up to the linear order. 
The deference between these two parametrizations emerges in the non-linear order: 
\begin{align}
\kappa w&=\kappa\Phi-\kappa^2\Phi^2-\frac{1}{16}\kappa^2\Psi_{\rho\sigma}\Psi^{\rho\sigma}+\cdots, \label{difference}\\
\kappa h_{\mu\nu}&=\kappa\Psi_{\mu\nu}-2\kappa^2\Phi\Psi_{\mu\nu}-\frac{1}{2}\kappa^2\Psi_\mu^{\ \rho}\Psi_{\rho\nu}+\frac{1}{8}\kappa^2\Psi_{\rho\sigma}\Psi^{\rho\sigma}\eta_{\mu\nu}+\cdots. \notag
\end{align} 

In the next section, we clarify how the parametrization difference contributes to the local dynamics of matter fields. 
Before the discussion, let us derive the gravitational propagator. 
In order to fix the gauge with respect to general coordinate invariance, we adopt the following gauge fixing term \cite{Tsamis1992}: 
\begin{align}
\mathcal{L}_\text{GF}&=-\frac{1}{2}a^{2}F_\mu F^\mu, \label{GF}\\
F_\mu&=\partial_\rho h_\mu^{\ \rho}-2\partial_\mu w+2h_\mu^{\ \rho}\partial_\rho\log a+4w\partial_\mu\log a. \notag
\end{align}
Note that in this paper, the Lagrangian density is defined including $\sqrt{-g}$ 
and the Lorentz indexes are raised and lowered by $\eta^{\mu\nu}$ and $\eta_{\mu\nu}$ respectively. 
The corresponding ghost term at the quadratic level is
\begin{align}
\mathcal{L}_\text{ghost}=&-a^2\partial^\nu\bar{b}^\mu 
\big\{{\eta}_{\mu\rho}\partial_\nu+\eta_{\nu\rho}\partial_\mu+2\eta_{\mu\nu}\partial_\rho(\log a)\big\}b^\rho\label{ghost}\\
&+\partial_\mu(a^2\bar{b}^\mu)
\big\{\partial_\nu+4\partial_\nu(\log a)\big\}b^\nu, \notag
\end{align}
where $b^\mu$ is the ghost field and $\bar{b}^\mu$ is the anti-ghost field. 
As far as we adopt the same gauge
\begin{align}
F_\mu=\partial_\rho \Psi_\mu^{\ \rho}-2\partial_\mu \Phi+2\Psi_\mu^{\ \rho}\partial_\rho\log a+4\Phi\partial_\mu\log a, 
\end{align}
we have only to identify the field components to obtain the propagator in the parametrization (\ref{Wpara1})-(\ref{Wpara2}): 
\begin{align}
w\to\Phi,\hspace{1em}h_{\mu\nu}\to\Psi_{\mu\nu}. 
\label{rename}\end{align}

In the parametrization (\ref{para1})-(\ref{para3}), the scalar density and the Ricci scalar are written as 
\begin{align}
\sqrt{-g}=\Omega^4,\hspace{1em}
R=\Omega^{-2}\tilde{R}-6\Omega^{-3}\tilde{g}^{\mu\nu}\nabla_\mu\partial_\nu\Omega, 
\label{com1,2}\end{align}
where $\tilde{R}$ is the Ricci scalar constructed from $\tilde{g}_{\mu\nu}$
\begin{align}
\tilde{R}=-\partial_\mu\partial_\nu\tilde{g}^{\mu\nu}
-\frac{1}{4}\tilde{g}^{\mu\nu}\tilde{g}^{\rho\sigma}\tilde{g}^{\alpha\beta}\partial_\mu\tilde{g}_{\rho\alpha}\partial_\nu\tilde{g}_{\sigma\beta}
+\frac{1}{2}\tilde{g}^{\mu\nu}\tilde{g}^{\rho\sigma}\tilde{g}^{\alpha\beta}\partial_\mu\tilde{g}_{\sigma\alpha}\partial_\rho\tilde{g}_{\nu\beta}. 
\label{com3}\end{align}
By substituting (\ref{com1,2}) and using the partial integration, the gravitational Lagrangian is
\begin{align}
\mathcal{L}_\text{gravity}
=\frac{1}{\kappa^2}\sqrt{-g}\big[R-2\Lambda\big]
=\frac{1}{\kappa^2}\big[\Omega^2\tilde{R}+6\tilde{g}^{\mu\nu}\partial_\mu\Omega\partial_\nu\Omega-6H^2\Omega^4\big], 
 \label{gravity}\end{align}
where $\Lambda=3H^2$. 

From (\ref{GF}), (\ref{ghost}), (\ref{com3}) and (\ref{gravity}), the quadratic part of the total gravitational Lagrangian density is
\begin{align}
\mathcal{L}_\text{quadratic}=a^4&\big[\ \frac{1}{2}a^{-2}\partial_\mu X\partial^\mu X-\frac{1}{4}a^{-2}\partial_\mu\tilde{h}^i_{\ j}\partial^\mu\tilde{h}^j_{\ i}
-a^{-2}\partial_\mu\bar{b}^i\partial^\mu b^i\label{quadratic}\\
&+\frac{1}{2}a^{-2}\partial_\mu h^{0i}\partial^\mu h^{0i}+H^2h^{0i}h^{0i}
-\frac{1}{2}a^{-2}\partial_\mu Y\partial^\mu Y-H^2Y^2\notag\\
&+a^{-2}\partial_\mu\bar{b}^0\partial^\mu b^0+2H^2\bar{b}^0b^0\big]. \notag
\end{align}
Here we have decomposed $h^i_{\ j},\ i,j=1, \cdots, 3$ into the trace and traceless part
\begin{align}
h^i_{\ j}=\tilde{h}^i_{\ j}+\frac{1}{3}h^k_{\ k}\delta^i_{\ j}=\tilde{h}^i_{\ j}+\frac{1}{3}h^{00}\delta^i_{\ j}.
\end{align} 
The action has been diagonalized by the following linear combination 
\begin{align}
X=2\sqrt{3}w-\frac{1}{\sqrt{3}}h^{00},\hspace{1em}Y=h^{00}-2w. 
\label{diagonalize}\end{align}
The quadratic action (\ref{quadratic}) contains two types of fields, 
massless and minimally coupled fields: $X,h^i_{\ j},b^i,\bar{b}^i$ 
and massless conformally coupled fields: $h^{0i},b^0,\bar{b}^0$, $Y$. 
We list the corresponding propagators as follows 
\begin{align}
\langle X(x)X(x')\rangle&=-\langle\varphi(x)\varphi(x')\rangle, \label{minimally}\\
\langle\tilde{h}^i_{\ j}(x)\tilde{h}^k_{\ l}(x')\rangle&=(\delta^{ik}\delta_{jl}+\delta^i_{\ l}\delta_j^{\ k}-\frac{2}{3}\delta^i_{\ j}\delta^k_{\ l})\langle\varphi(x)\varphi(x')\rangle, \notag\\
\langle b^i(x)\bar{b}^j(x')\rangle&=\delta^{ij}\langle\varphi(x)\varphi(x')\rangle, \notag
\end{align}
\begin{align}
\langle h^{0i}(x)h^{0j}(x')\rangle&=-\delta^{ij}\langle\phi(x)\phi(x')\rangle, \label{conformally}\\
\langle Y(x)Y(x')\rangle&=\langle\phi(x)\phi(x')\rangle, \notag\\
\langle b^0(x)\bar{b}^0(x')\rangle&=-\langle\phi(x)\phi(x')\rangle. \notag
\end{align}

Here $\varphi$ denotes a massless and minimally coupled scalar field and $\phi$ denotes a massless conformally coupled scalar field
\begin{align}
\langle\varphi(x)\varphi(x')\rangle&=\frac{H^2}{4\pi^2}\big\{\frac{1}{y}-\frac{1}{2}\log y+\frac{1}{2}\log a(\tau)a(\tau')+1-\gamma\big\}, 
\label{minimally0}\end{align} 
\begin{align}
\langle\phi(x)\phi(x')\rangle&=\frac{H^2}{4\pi^2}\frac{1}{y}, 
\label{conformally0}\end{align}
where $\gamma$ is Euler's constant and $y$ is the dS invariant distance
\begin{align}
y=\Delta x_\mu\Delta x^\mu/\tau\tau',\hspace{1em}\Delta x^0=\tau-\tau',\ \Delta x^i=x^i-{x'}^i. 
\end{align}
The existence of  the logarithmic term: $\log a(\tau)a(\tau')$ indicates that 
the propagator for a massless and minimally coupled scalar field breaks the dS symmetry. 
In particular, it breaks the scale invariance
\begin{align}
\tau\to C\tau,\hspace{1em}x^i\to Cx^i. 
\label{scale}\end{align}

To explain what causes the dS symmetry breaking, 
we recall the wave function for a massless and minimally coupled field
\begin{align}
\phi_{\bf p}(x)=\frac{H\tau}{\sqrt{2p}}(1-i\frac{1}{p\tau})e^{-ip\tau+i{\bf p}\cdot{\bf x}}. 
\end{align}
At the sub-horizon scale $P\equiv p/a(\tau)\gg H \Leftrightarrow p|\tau|\gg 1$, 
this wave function approaches to that in Minkowski space up to a cosmic scale factor
\begin{align}
\phi_{\bf p}(x)\sim\frac{H\tau}{\sqrt{2p}}e^{-ip\tau+i{\bf p}\cdot{\bf x}}. 
\end{align}
On the other hand, the behavior at the super-horizon scale $P\ll H$ is
\begin{align}
\phi_{\bf p}(x)\sim\frac{H}{\sqrt{2p^3}}e^{i{\bf p}\cdot{\bf x}}. 
\end{align}
The IR behavior indicates that the corresponding propagator has a scale invariant spectrum. 
As a direct consequence of it, the propagator has a logarithmic divergence from the IR contributions in the infinite volume limit. 

To regularize the IR divergence, we introduce an IR cut-off $\varepsilon_0$ which fixes the minimum value of the comoving momentum. 
The minimum value of the physical momentum is $\varepsilon_0/a(\tau )$ as their wavelength is stretched by cosmic expansion.
With this prescription, more degrees of freedom accumulate at the super-horizon scale with cosmic evolution. 
Due to this increase, the propagator acquires the growing time dependence which spoils the dS symmetry. 
In tribute to its origin, we call this type of dS symmetry breaking term the IR logarithm. 
Physically speaking, $1/\varepsilon_0$ is recognized as an initial size of universe when the exponential expansion starts. 
For simplicity, we set $\varepsilon_0=H$ in (\ref{minimally0}). 

As there is explicit time dependence in the propagator, 
physical quantities can acquire time dependence through the quantum loop corrections. 
We call them the quantum IR effects in dS space. 
Focusing on the leading IR effects, we introduce an approximation. 
We can neglect conformally coupled modes of gravity since they do not induce the IR logarithm. 
Then, the following two modes are identified as 
\begin{align}
h^{00}\simeq 2w\simeq\frac{\sqrt{3}}{2}X. 
\label{diagonalize1}\end{align}
From (\ref{minimally}) and (\ref{diagonalize1}), 
we have only to focus on the following propagators after retaining massless and minimally coupled modes from gravity
\begin{align}
\langle h^{00}(x)h^{00}(x')\rangle&\simeq-\frac{3}{4}\langle\varphi(x)\varphi(x')\rangle, \label{gravityp}\\
\langle h^{00}(x)h^i_{\ j}(x')\rangle&\simeq-\frac{1}{4}\delta^i_{\ j}\langle\varphi(x)\varphi(x')\rangle, \notag\\
\langle h^i_{\ j}(x)h^k_{\ l}(x')\rangle&\simeq(\delta^{ik}\delta_{jl}+\delta^i_{\ l}\delta_j^{\ k}-\frac{3}{4}\delta^i_{\ j}\delta^k_{\ l})\langle\varphi(x)\varphi(x')\rangle. \notag
\end{align}

\section{Quantization scheme dependence}
\setcounter{equation}{0}

In this section, we clarify the quantization scheme dependence of soft gravitational effects on the dS background. 
Specifically, we investigate the IR effects on the local dynamics of matter fields at the sub-horizon scale which are directly observable. 
Let us consider a matter action with conformal invariance first such as the massless Dirac field
\begin{align}
\mathcal{L}_D=i\sqrt{-g}\bar{\psi}e^{\mu}_{\ a}\gamma^a \nabla_\mu\psi, 
\label{kD}\end{align}
where $e^\mu_{\ a}$ is the vierbein, $\nabla_\mu$ is the covariant derivative and $\gamma^a$ 
satisfy
\begin{align}
\gamma^a\gamma^b+\gamma^b\gamma^a=-2\eta^{ab}.  
\end{align} 
Since dS space is conformally flat, the matter action possesses the symmetry of flat space, namely Lorentz invariance at the classical level. 
In \cite{Miao2012}, the IR effects on a free Dirac field theory with a small mass relative to the Hubble scale have been investigated. 
We emphasize that in this paper, we can neglect the mass term because we focus on the local dynamics at the sub-horizon scale.  
In the same way, our investigation covers a more generic situation such as the massless and minimally coupled scalar field 
\begin{align}
\mathcal{L}_s=-\frac{1}{2}\sqrt{-g}g^{\mu\nu}\partial_\mu\varphi\partial_\nu\varphi, 
\label{ks}\end{align}
where conformal invariance of a matter action holds at short distance and Lorentz invariance appears as an effective symmetry. 
Considering quantum corrections on them, it is a non-trivial question whether soft gravitons preserve the effective Lorentz invariance of the local dynamics. 
In order to answer this question, we evaluate the IR effects on the kinetic terms (\ref{kD}), (\ref{ks}). 

Let us consider the field redefinition by the conformal transformation. 
We should note that there is a difference of the matter field redefinition between \cite{KitamotoSD,KitamotoG} 
and \cite{Kahya2007,Miao2005(1),Miao2005(2),Miao2005(3)}. 
The following field redefinition is adopted in \cite{KitamotoSD,KitamotoG}: 
\begin{align}
\varphi_0=\Omega\varphi,\hspace{1em}\psi_0=\Omega^{\frac{3}{2}}\psi. 
\label{paraf1}\end{align}
The corresponding Lagrangians are
\begin{align}
\mathcal{L}_s=-\frac{1}{2}\tilde{g}^{\mu\nu}\partial_\mu\varphi_0\partial_\nu\varphi_0
-\frac{1}{2}\Omega^{-1}\partial_\mu\big\{\tilde{g}^{\mu\nu}\partial_\nu\Omega\big\}\varphi^2_0, 
\label{paraf2}\end{align}
\begin{align}
\mathcal{L}_D=i\bar{\psi}_0\gamma^a \tilde{e}^{\mu}_{\ a}\nabla_\mu|_{\tilde{g}_{\rho\sigma}}\psi_0, 
\label{paraf3}\end{align}
where $\tilde{e}^\mu_{\ a}$ is 
\begin{align}
\tilde{e}^\mu_{\ a}=(e^{-\frac{1}{2}\kappa h})^\mu_{\ a}, 
\end{align}
and $\nabla_\mu|_{\tilde{g}_{\rho\sigma}}$ denotes the covariant derivative with respect to $\tilde{g}_{\rho\sigma}$. 

On the other hand, the following field redefinition is adopted in \cite{Kahya2007,Miao2005(1),Miao2005(2),Miao2005(3)}: 
\begin{align}
\phi_w=a\phi,\hspace{1em}\psi_w=a^\frac{3}{2}\psi. 
\label{Wparaf1}\end{align}
The corresponding Lagrangians are
\begin{align}
\mathcal{L}_s=-\frac{1}{2}e^{2\kappa w}\tilde{g}^{\mu\nu}\partial_\mu\varphi_w\partial_\nu\varphi_w
-\frac{1}{2}a^{-1}\partial_\mu\big\{e^{2\kappa w}\tilde{g}^{\mu\nu}\partial_\nu a\big\}\varphi^2_w, 
\label{Wparaf2}\end{align}
\begin{align}
\mathcal{L}_D=i\bar{\psi}_w\gamma^a e^{3\kappa w}\tilde{e}^{\mu}_{\ a}\nabla_\mu|_{e^{2\kappa w}\tilde{g}_{\rho\sigma}}\psi_w. 
\label{Wparaf3}\end{align}

First, we review the IR effects in the parametrization (\ref{para1})-(\ref{para3}), with the field redefinition (\ref{paraf1}). 
In investigating interacting field theories on a time dependent background like dS space, 
we need to adopt the Schwinger-Keldysh path \cite{Schwinger1961,Keldysh1964} 
\begin{align}
&\parbox{\SKpathl}{\usebox{\SKpath}}, \label{SKpath}\\
&\hspace{3em}\int_C dt = \int^\infty_{-\infty} dt_+ - \int^\infty_{-\infty} dt_-. \notag
\end{align}
Since there are two time indices $+,-$ in this path, the propagator has four components 
\begin{align}
\begin{pmatrix} \langle\varphi_+(x)\varphi_+(x')\rangle & \langle\varphi_+(x)\varphi_-(x')\rangle \\
\langle\varphi_-(x)\varphi_+(x')\rangle & \langle\varphi_-(x)\varphi_-(x')\rangle \end{pmatrix}
=\begin{pmatrix} \langle T\varphi(x)\varphi(x')\rangle & \langle\varphi(x')\varphi(x)\rangle \\
\langle\varphi(x)\varphi(x')\rangle & \langle\tilde{T}\varphi(x)\varphi(x')\rangle \end{pmatrix}, 
\label{4propagators}\end{align}
where $\varphi$ denotes the quantum fluctuation of an arbitrary field component and $\tilde{T}$ denotes the anti-time ordering. 
We divide the field component into the classical field and the quantum fluctuation
\begin{align}
\varphi\to\hat{\varphi}+\varphi. 
\end{align}
By ordering the quantum fluctuation along the path (\ref{SKpath}), 
we can derive the effective equation of motion \cite{Hu1997} 
\begin{align}
\frac{\delta\Gamma[\hat{\varphi}_+,\hat{\varphi}_-]}{\delta\hat{\varphi}_+}\big|_{\hat{\varphi}_+=\hat{\varphi}_-=\hat{\varphi}}=0. 
\end{align}
Here $\Gamma$ denotes the effective action. 
Since our models contain massless field components, we need to set the classical fields to be off-shell. 
As explained later, the IR singularity at the on-shell limit is irrelevant with the dS symmetry breaking. 

Concerning the internal loop contributions at the one-loop level, 
we can clearly separate IR contributions and  ultra-violet (UV) contributions relative to the Hubble scale. 
As far as IR logarithms are concerned, we can safely ignore UV contributions and hence UV divergences altogether. 
This is the strategy we adopt in this paper. 
UV divergences and the counter terms never depend on time explicitly as short distance degrees of freedom is held constant.

Up to the one-loop level, the effective equations of motion are written as
\begin{align}
&\partial^2\hat{\varphi}_0(x)
+\frac{1}{2}\kappa^2\partial_\mu\{\langle (h^{\mu\rho})_+(x)(h_\rho^{\ \nu})_+(x)\rangle\partial_\nu\hat{\varphi}_0(x)\}\label{sEoM1}\\
&+i\kappa^2\partial_\mu\int d^4x'\ c_{AB}\partial_\sigma'
\big\{\langle (h^{\mu\nu})_+(x)(h^{\rho\sigma})_A(x')\rangle\langle\partial_\nu(\varphi_0)_+(x)\partial_\rho'(\varphi_0)_B(x')\rangle\big\}\hat{\varphi}_0(x') \notag\\
&+i\kappa^2\int d^4x'\ c_{AB}\langle \partial^2w_+(x){\partial'}^2w_A(x')\rangle\langle(\varphi_0)_+(x)(\varphi_0)_B(x')\rangle\hat{\varphi}_0(x')
\simeq 0, \notag
\end{align}
\begin{align}
&i\gamma^\mu\partial_\mu\hat{\psi}_0(x)
+i\frac{\kappa^2}{8}\langle (h^\mu_{\ \rho})_+(x)(h^\rho_{\ a})_+(x)\rangle\gamma^a\partial_\mu\hat{\psi}_0(x)\label{DEoM1}\\
&+i\frac{\kappa^2}{4}\int d^4x'\ c_{AB}\partial_\nu'
\big\{\langle (h^\mu_{\ a})_+(x)(h^\nu_{\ b})_A(x')\rangle\gamma^a\langle\partial_\mu(\psi_0)_+(x)(\bar{\psi}_0)_B(x')\rangle\big\}\gamma^b\hat{\psi}_0(x')
\simeq 0, \notag
\end{align} 
where $c_{AB}$ is identified as 
\begin{align}
c_{AB}=\begin{pmatrix} 1 & 0 \\ 0 & -1\end{pmatrix}. 
\end{align}

In (\ref{sEoM1}) and (\ref{DEoM1}), 
we have only shown the terms with the IR logarithms which are of the following type: $\log a(\tau)\partial\partial\hat{\varphi}_0(x)$, $\log a(\tau)\partial\hat{\psi}_0(x)$. 
At the sub-horizon scale, we can neglect the derivative of the scale factor $\partial_\mu a$ in comparison to the external momentum of the matter: 
\begin{align}
P\gg H\ \Leftrightarrow\ 
a\partial_\mu \hat{\varphi}_0\gg (\partial_\mu a)\hat{\varphi}_0,\ a\partial_\mu \hat{\psi}_0\gg (\partial_\mu a)\hat{\psi}_0,  
\label{sub}\end{align}
where $P$ denotes the external physical momentum scale of the matter. 
It implies that the dS background evolves slowly in comparison to the matter dynamics at the sub-horizon scale.
Furthermore, we have assumed that the differentiated gravitons do not induce the IR logarithms. 
Since the four-point vertices contribute only to the local dynamics, that can be easily confirmed for them. 
Specifically the contributions from the four-point vertices are evaluated as 
\begin{align}
\frac{1}{2}\kappa^2\partial_\mu\big\{\langle (h^{\mu\rho})_+(x)(h_\rho^{\ \nu})_+(x)\rangle\partial_\nu\hat{\varphi}_0(x)\big\} 
\simeq\frac{\kappa^2H^2}{4\pi^2}\log a(\tau)\big\{\frac{3}{8}\partial_0^2+\frac{13}{8}\partial_i^2\big\}\hat{\varphi}_0(x), 
\label{sEoM2}\end{align}
\begin{align}
i\frac{\kappa^2}{8}\langle (h^\mu_{\ \rho})_+(x)(h^\rho_{\ a})_+(x)\rangle\gamma^a\partial_\mu\hat{\psi}_0(x) 
\simeq i\frac{\kappa^2H^2}{4\pi^2}\log a(\tau)\big\{-\frac{3}{32}\gamma^0\partial_0+\frac{13}{32}\gamma^i\partial_i\big\}\hat{\psi}_0(x). 
\label{DEoM2}\end{align}
Here we have used the approximate identities (\ref{gravityp}) and extracted the $\mathcal{O}(\log a(\tau))$ terms. 

On the other hand, the three-point vertices lead to the local and non-local contributions. 
In order to extract the local dynamics with the IR logarithm, it is useful to keep the following points in mind. 
As for the classical fields, we need to expand them up to the following orders: 
\begin{align}
&\hat{\varphi}_0(x')\to (1-\Delta x^\alpha\partial_\alpha+\frac{1}{2}\Delta x^\alpha\Delta x^\beta \partial_\alpha\partial_\beta)\hat{\varphi}_0(x), \label{expansion}\\
&\hat{\psi}_0(x')\to (1-\Delta x^\alpha\partial_\alpha)\hat{\psi}_0(x). \notag
\end{align}
As for the quantum fluctuations, we claim that 
the IR logarithms come only from the propagator of a massless and minimally coupled field left intact by differential operators as we explain subsequently.
To evaluate the integrals up to $\mathcal{O}(\log a(\tau))$, we may extract the dS broken term in the propagator 
\begin{align}
\langle\varphi_0(x)\varphi_0(x')\rangle,\langle h^{\mu\nu}(x)h^{\rho\sigma}(x)\rangle\to \frac{H^2}{8\pi^2}\log a(\tau)a(\tau'). 
\label{IR}\end{align}
and move it out of the integrals: 
\begin{align}
&\ i\kappa^2\partial_\mu\int d^4x'\ c_{AB}\partial_\sigma'\big\{\langle (h^{\mu\nu})_+(x)(h^{\rho\sigma})_A(x')\rangle
\langle\partial_\nu(\varphi_0)_+(x)\partial_\rho'(\varphi_0)_B(x')\rangle\big\}\hat{\varphi}_0(x') \label{int1}\\
\simeq&\ i\kappa^2\langle (h^{\mu\nu})_+(x)(h^{\rho\sigma})_+(x)\rangle
\partial_\mu\int d^4x'\ c_{A}
\langle\partial_\nu(\varphi_0)_+(x)\partial_\rho'\partial_\sigma'(\varphi_0)_A(x')\rangle\hat{\varphi}_0(x'), \notag
\end{align}
\begin{align}
&\ i\kappa^2\int d^4x'\ c_{AB}\langle\partial^2w_+(x){\partial'}^2w_A(x')\rangle
\langle(\varphi_0)_+(x)(\varphi_0)_B(x')\rangle\hat{\varphi}_0(x') \label{int2}\\
\simeq&\ i\kappa^2\langle(\varphi_0)_+(x)(\varphi_0)_+(x)\rangle
\int d^4x'\ c_{A}\langle \partial^2w_+(x){\partial'}^2w_A(x')\rangle\hat{\varphi}_0(x'), \notag
\end{align}
\begin{align}
&\ i\frac{\kappa^2}{4}\int d^4x'\ c_{AB}\partial_\nu'\big\{\langle (h^\mu_{\ a})_+(x)(h^\nu_{\ b})_A(x')\rangle
\gamma^a\langle\partial_\mu(\psi_0)_+(x)(\bar{\psi}_0)_B(x')\rangle\big\}\gamma^b\hat{\psi}_0(x') \label{int3}\\
\simeq&\ i\frac{\kappa^2}{4}\langle (h^\mu_{\ a})_+(x)(h^\nu_{\ b})_+(x)\rangle
\int d^4x'\ c_{A}\gamma^a\langle\partial_\mu(\psi_0)_+(x)\partial_\nu'(\bar{\psi}_0)_A(x')\rangle\gamma^b\hat{\psi}_0(x'), \notag
\end{align}
where $c_A$ is identified as $c_\pm=\pm 1$ (the order of double-sign corresponds). 
This approximation method has been introduced in Yukawa theory and scalar QED \cite{Duffy2005,Woodard2006,Prokopec2007}. 

The dS invariant terms which we have neglected in (\ref{IR}) seem to induce the IR logarithms after the time integration at first sight
\begin{align}
\int^\tau_{\tau_i}\frac{d\tau'}{\tau'}, 
\end{align}
where $\tau_i$ denotes an initial time which cuts-off the IR logarithms. 
However if we set the classical fields to be off-shell, the integrand is constantly oscillating. 
So the integral over the negatively large conformal time is suppressed by the inverse of the virtuality. 

At the on-shell limit, we can argue that this kind of IR divergences cancel just like Kinoshita-Lee-Nauenberg theorem in field theory \cite{Kinoshita1962,Lee1964}. 
We need to recall that a matter particle is accompanied by soft gravitons. 
We cannot distinguish a single particle state and a state with a particle and a soft or collinear graviton if the difference of the total energy is smaller than the energy resolution for a fixed total momentum. 
In fact, (\ref{IR}) is not entirely correct as we need to include such an off-shell contribution in (\ref{IR}). 
Specifically we claim that the non-local contributions from the following combination are IR finite: 
\begin{align}
\parbox{\ThreeAl}{\usebox{\ThreeA}},\hspace{2em}\parbox{\ThreeBl}{\usebox{\ThreeB}}. 
\end{align}
Here the vertical line segment indicates a derivative with respect to the classical field 
and propagators cut by the dotted line are on-shell. 
The diagram $A$ contains the local and non-local terms with IR logarithms. 
The diagram $B$ represents the squared amplitude $\phi\to h^{\mu\nu}\phi$ with the classical field and contains the non-local terms with IR logarithms. 
In a derivation of a Boltzmann equation, the cancellation of the non-local IR singularities follows from the Schwinger-Dyson equation. 
We have indeed found that the leading IR effects are canceled between the real and virtual processes in dS space \cite{Kitamoto2010}. 
The investigation has been performed in $\varphi^3$, $\varphi^4$ theories. 
Nevertheless we can argue that the cancellation takes place in any unitary model as the total spectral weight is preserved.
 
As a consequence, the lower bound of the integral is given by the virtuality $\sqrt{p_\mu p^\mu}$ or the energy resolution of the observation $\Delta \epsilon$:
\begin{align}
\int^\tau_{-1/\sqrt{p_\mu p^\mu}}\frac{d\tau'}{\tau'}=\log(-\sqrt{p_\mu p^\mu}\tau),\hspace{1em}
\int^\tau_{-1/\Delta\epsilon}\frac{d\tau'}{\tau'}=\log(-\Delta\epsilon\tau). 
\label{bound}\end{align}
These integrals are invariant under the scale transformation (\ref{scale}). 
In other words, these logarithms are time independent when they are expressed by the physical momentum scales $\sqrt{P_\mu P^\mu}=-\sqrt{p_\mu p^\mu}H\tau$, $\Delta E=-\Delta \epsilon H\tau$. 
That is why we have neglected them. 
In order to probe local dynamics of matter fields, we can perform experiments in the laboratory. 
We thus hold the physical momentum fixed in such a situation. 

As stated in (\ref{sub}), we focus on the coefficients of $\partial\partial\hat{\varphi}_0(x)$, $\partial\hat{\psi}_0(x)$. 
In the remaining integrals, 
the twice differentiated propagators contribute to the local terms as follows
\begin{align}
\partial_\mu\partial_\nu\frac{1}{\Delta x_{++}^2}\to -4i\pi^2\delta_\mu^{\ 0}\delta_\nu^{\ 0}\delta^{(4)}(x-x'). 
\end{align}
So to investigate the local dynamics, 
we have only to focus on the most singular parts of the propagators 
\begin{align}
&\langle\phi_0(x)\phi_0(x')\rangle=\frac{1}{4\pi^2}\frac{1}{\Delta x^2}, \hspace{1em}
\langle\psi_0(x)\bar{\psi}_0(x')\rangle=i\gamma^\rho\partial_\rho\langle\phi_0(x)\phi_0(x')\rangle, \label{flat}\\
&\langle\varphi_0(x)\varphi_0(x')\rangle\to \frac{1}{4\pi^2}\frac{1}{\Delta x^2}, \hspace{1em}
\langle w(x)w(x')\rangle\to -\frac{1}{8}\times\frac{H^2}{4\pi^2}\frac{1}{y}. \notag
\end{align} 
From (\ref{expansion}) and (\ref{flat}), the remaining integrals are evaluated as 
\begin{align}
&i\partial_\mu\int d^4x'\ c_{A}\langle\partial_\nu(\varphi_0)_+(x)\partial_\rho'\partial_\sigma'(\varphi_0)_A(x')\rangle\hat{\varphi}_0(x') \label{int4}\\
\to\big\{&\delta_\mu^{\ 0}\delta_\nu^{\ 0}\partial_\rho\partial_\sigma+\delta_\mu^{\ 0}\delta_\rho^{\ 0}\partial_\nu\partial_\sigma+\delta_\mu^{\ 0}\delta_\sigma^{\ 0}\partial_\nu\partial_\rho
+\delta_\nu^{\ 0}\delta_\rho^{\ 0}\partial_\mu\partial_\sigma+\delta_\nu^{\ 0}\delta_\sigma^{\ 0}\partial_\mu\partial_\rho+\delta_\rho^{\ 0}\delta_\sigma^{\ 0}\partial_\mu\partial_\nu\notag\\
&-2(\delta_\mu^{\ 0}\delta_\nu^{\ 0}\delta_\rho^{\ 0}\partial_\sigma+\delta_\mu^{\ 0}\delta_\nu^{\ 0}\delta_\sigma^{\ 0}\partial_\rho
+\delta_\mu^{\ 0}\delta_\rho^{\ 0}\delta_\sigma^{\ 0}\partial_\nu+\delta_\nu^{\ 0}\delta_\rho^{\ 0}\delta_\sigma^{\ 0}\partial_\mu)\partial_0\notag\\
&+4\delta_\mu^{\ 0}\delta_\nu^{\ 0}\delta_\rho^{\ 0}\delta_\sigma^{\ 0}\partial_0^2+\delta_\mu^{\ 0}\delta_\nu^{\ 0}\delta_\rho^{\ 0}\delta_\sigma^{\ 0}\partial^2
\big\}\hat{\varphi}_0(x), \notag
\end{align}
\begin{align}
i\int d^4x'\ c_{A}\langle \partial^2w_+(x){\partial'}^2w_A(x')\rangle\hat{\varphi}_0(x')\to\frac{1}{8}a^{-2}(\tau)\partial^2\hat{\varphi}_0(x), 
\label{int5}\end{align}
\begin{align}
&\ i\int d^4x'\ c_{A}\gamma^a\langle\partial_\mu(\psi_0)_+(x)\partial_\nu'(\bar{\psi}_0)_A(x')\rangle\gamma^b\hat{\psi}_0(x') \label{int6}\\
\to&\ i\gamma^a\gamma^\rho\gamma^b
\big\{-(\delta_\mu^{\ 0}\delta_\nu^{\ 0}\partial_\rho+\delta_\mu^{\ 0}\delta_\rho^{\ 0}\partial_\nu+\delta_\nu^{\ 0}\delta_\rho^{\ 0}\partial_\mu)
+2\delta_\mu^{\ 0}\delta_\nu^{\ 0}\delta_\rho^{\ 0}\partial_0\big\}\hat{\psi}_0(x). \notag
\end{align}
For more details of the derivation, please refer to \cite{KitamotoSD}. 
From (\ref{gravityp}), (\ref{int1})-(\ref{int3}) and (\ref{int4})-(\ref{int6}), 
the local terms from the three-point vertices are evaluated up to $\mathcal{O}(\log a(\tau))$ as follows 
\begin{align}
&\ i\kappa^2\partial_\mu\int d^4x'\ c_{AB}
\partial_\sigma'\big\{\langle (h^{\mu\nu})_+(x)(h^{\rho\sigma})_A(x')\rangle\langle\partial_\nu(\varphi_0)_+(x)\partial_\rho'(\varphi_0)_B(x')\rangle\big\}\hat{\varphi}_0(x') \label{sEoM31}\\
\to&\ \frac{\kappa^2H^2}{4\pi^2}\log a(\tau)\big\{-\frac{3}{4}\partial_0^2-\frac{5}{4}\partial_i^2\big\}\hat{\varphi}_0(x), \notag
\end{align}
\begin{align}
&\ i\kappa^2\int d^4x'\ c_{AB}\langle \partial^2w_+(x){\partial'}^2w_A(x')\rangle
\langle(\varphi_0)_+(x)(\varphi_0)_B(x')\rangle\hat{\varphi}_0(x') \label{sEoM32}\\
\to&\ \frac{\kappa^2H^2}{4\pi^2}\log a(\tau)\times \frac{1}{8}\partial^2\hat{\varphi}_0(x), \notag
\end{align}
\begin{align}
&\ i\frac{\kappa^2}{4}\int d^4x'\ c_{AB}\partial_\nu'\big\{\langle (h^\mu_{\ a})_+(x)(h^\nu_{\ b})_A(x')\rangle
\gamma^a\langle\partial_\mu(\psi_0)_+(x)(\bar{\psi}_0)_B(x')\rangle\big\}\gamma^b\hat{\psi}_0(x') \label{DEoM3}\\
\to&\ i\frac{\kappa^2H^2}{4\pi^2}\log a(\tau)\big\{\frac{3}{16}\gamma^0\partial_0-\frac{5}{16}\gamma^i\partial_i\big\}\hat{\psi}_0(x). \notag
\end{align} 

From (\ref{sEoM2})-(\ref{DEoM2}) and (\ref{sEoM31})-(\ref{DEoM3}), the effective equations of motion up to the one-loop level are
\begin{align}
\big\{\partial^2+\frac{1}{2}\frac{\kappa^2H^2}{4\pi^2}\log a(\tau)\partial^2\big\}\hat{\varphi}_0(x)\simeq 0, 
\label{sEoM4}\end{align}
\begin{align}
i\big\{\gamma^\mu\partial_\mu+\frac{3}{32}\frac{\kappa^2H^2}{4\pi^2}\log a(\tau)\gamma^\mu\partial_\mu\big\}\hat{\psi}_0(x)\simeq 0. 
\label{DEoM4}\end{align}
The relative weights between the time derivative terms and the spatial derivative terms are equal to $-1$. 
In this regard, the total of the IR effects preserve the effective Lorentz invariance while the IR effect from each diagram does not always preserve it. 
Let us recall the left-hand side of (\ref{sEoM4}) and (\ref{DEoM4}) are the variations of the off-shell effective action with respect to fields. 
We apply a renormalization procedure to the off-shell effective action to make the kinetic terms to be canonical. 
In this way the final IR logarithmic effects can be absorbed by the wave function renormalization factors 
\begin{align}
\varphi_0\to Z_\varphi\varphi_0,\hspace{1em}Z_\varphi\simeq 1-\frac{1}{4}\frac{\kappa^2H^2}{4\pi^2}\log a(\tau), 
\end{align}
\begin{align}
\psi_0\to Z_\psi\psi_0,\hspace{1em}Z_\psi\simeq 1-\frac{3}{64}\frac{\kappa^2H^2}{4\pi^2}\log a(\tau).  
\end{align}
Note that we can neglect the derivative of $\log a(\tau)$ at the sub-horizon scale. 
We conclude that there is no physical effect from IR logarithms in free field theories at the sub-horizon scale. 

Next, let us translate (\ref{sEoM4}), (\ref{DEoM4}) into the quantum equations 
in the parametrization (\ref{Wpara1})-(\ref{Wpara2}), with the field redefinition (\ref{Wparaf1}). 
The parametrization difference of the metric (\ref{difference}) contributes only to the tadpole diagrams at the one-loop level: 
\begin{align}
\Delta(\delta\Gamma/\delta\hat{\varphi})|_\text{metric}=
-\kappa\partial_\mu\big\{\langle (h^{\mu\nu})_+(x)\rangle|_\text{NL}\partial_\nu\hat{\varphi}_0(x)\big\}, 
\label{Ps1}\end{align}
\begin{align}
\Delta(\delta\Gamma/\delta\hat{\psi})|_\text{metric}=
-i\frac{\kappa}{2}\langle (h^\mu_{\ a})_+(x)\rangle|_\text{NL}\gamma^a\partial_\mu\hat{\psi}_0(x), 
\label{PD1}\end{align}
where $\kappa\langle h_{\mu\nu}(x)\rangle|_\text{NL}$ is identified as
\begin{align}
\kappa\langle h_{\mu\nu}(x)\rangle|_\text{NL}
=-2\kappa^2\langle\Phi(x)\Psi_{\mu\nu}(x)\rangle
-\frac{1}{2}\kappa^2\langle\Psi_{\mu}^{\ \rho}(x)\Psi_{\rho\nu}(x)\rangle
+\frac{1}{8}\kappa^2\langle\Psi_{\rho\sigma}(x)\Psi^{\sigma\rho}(x)\rangle\eta_{\mu\nu}. 
\end{align}
From (\ref{rename}) and (\ref{gravityp}), these differences are evaluated as 
\begin{align}
\Delta(\delta\Gamma/\delta\hat{\varphi})|_\text{metric}
\simeq\frac{\kappa^2H^2}{4\pi^2}\log a(\tau)\big\{\frac{3}{4}\partial_0^2+\frac{1}{4}\partial_i^2\big\}\hat{\varphi}_0(x), 
\label{Ps2}\end{align}
\begin{align}
\Delta(\delta\Gamma/\delta\hat{\psi})|_\text{metric}\simeq
i\frac{\kappa^2H^2}{4\pi^2}\log a(\tau)\big\{-\frac{3}{8}\gamma^0\partial_0+\frac{1}{8}\gamma^i\partial_i\big\}\hat{\psi}_0(x). 
\label{PD2}\end{align}

In addition to them, 
the matter field redefinition $\varphi_0\to\varphi_w$, $\psi_0\to\psi_w$ does not only relabel the matter fields 
but contributes to the quantum equations as: 
\begin{align}
&\ \Delta(\delta\Gamma/\delta\hat{\varphi})|_\text{field} \label{Fs1}\\
=&\ \partial_\mu\big\{\big(2\kappa\langle w_+(x)\rangle|_\text{NL}\eta^{\mu\nu}
+2\kappa^2\langle w_+(x)w_+(x)\rangle\eta^{\mu\nu}-2\kappa^2\langle w_+(x)(h^{\mu\nu})_+(x)\rangle\big)
\partial_\nu\hat{\varphi}_w(x)\big\} \notag\\
&+i\kappa^2\partial_\mu\int d^4x'\ c_{AB}\partial_\sigma'
\big\{\big(4\langle w_+(x)w_A(x')\rangle\eta^{\mu\nu}\eta^{\rho\sigma}
-2\langle w_+(x)(h^{\rho\sigma})_A(x')\rangle\eta^{\mu\nu} \notag\\
&\hspace{10em}-2\langle (h^{\mu\nu})_+(x)w_A(x')\rangle\eta^{\rho\sigma}\big)
\langle\partial_\nu(\varphi_w)_+(x)\partial_\rho'(\varphi_w)_B(x')\rangle\big\}\hat{\varphi}_w(x') \notag\\
&-i\kappa^2\int d^4x'\ c_{AB}\langle \partial^2w_+(x){\partial'}^2w_A(x')\rangle\langle(\varphi_w)_+(x)(\varphi_w)_B(x')\rangle\hat{\varphi}_w(x'), \notag
\end{align}
\begin{align}
&\ \Delta(\delta\Gamma/\delta\hat{\psi})|_\text{field} \label{FD1}\\
=&\ i\big(3\kappa\langle w_+(x)\rangle|_\text{NL}\eta^\mu_{\ a}+\frac{9}{2}\kappa^2\langle w_+(x)w_+(x)\rangle\eta^\mu_{\ a}
-\frac{3}{2}\kappa^2\langle w_+(x)(h^\mu_{\ a})_+(x)\rangle\big)
\gamma^a\partial_\nu\hat{\psi}_w(x) \notag\\
&+i\kappa^2\int d^4x'\ c_{AB}\partial_\nu'\big\{ 
\big(9\langle w_+(x)w_A(x')\rangle\eta^\mu_{\ a}\eta^\nu_{\ b}-\frac{3}{2}\langle w_+(x)(h^\nu_{\ b})_A(x')\rangle\eta^\mu_{\ a}\notag\\
&\hspace{9em}-\frac{3}{2}\langle (h^\mu_{\ a})_+(x)w_A(x')\rangle\eta^\nu_{\ b}\big)
\gamma^a\langle\partial_\mu(\psi_w)_+(x)(\bar{\psi}_w)_B(x')\rangle\big\}\gamma^b\hat{\psi}_w(x'), \notag
\end{align}
where $\kappa\langle w(x)\rangle|_\text{NL}$ originates in the parametrization difference of the metric (\ref{difference})
\begin{align}
\kappa\langle w(x)\rangle|_\text{NL}=-\kappa^2\langle\Phi^2(x)\rangle-\frac{1}{16}\kappa^2\langle\Psi_{\rho\sigma}(x)\Psi^{\rho\sigma}(x)\rangle. 
\end{align}
The last line of (\ref{Fs1}) denotes the elimination of (\ref{sEoM32}) from (\ref{sEoM1}). 
In a similar way to (\ref{sEoM31})-(\ref{DEoM3}), we can evaluate each term of (\ref{Fs1}) and (\ref{FD1}).  
By summing them, the contributions from the different field redefinition manifest as 
\begin{align}
\Delta(\delta\Gamma/\delta\hat{\varphi})|_\text{field}
\simeq\frac{\kappa^2H^2}{4\pi^2}\log a(\tau)\big\{-\frac{1}{4}\partial_0^2-\frac{3}{4}\partial_i^2\big\}\hat{\varphi}_w(x), 
\label{Fs2}\end{align}
\begin{align}
\Delta(\delta\Gamma/\delta\hat{\psi})|_\text{field}\simeq
i\frac{\kappa^2H^2}{4\pi^2}\log a(\tau)\big\{\frac{9}{32}\gamma^0\partial_0-\frac{15}{32}\gamma^i\partial_i\big\}\hat{\psi}_w(x). 
\label{FD2}\end{align}
See Appendix {\ref{A:A}} for detailed calculations.

From (\ref{sEoM4})-(\ref{DEoM4}), (\ref{Ps2})-(\ref{PD2}) and (\ref{Fs2})-(\ref{FD2}), 
the effective equations of motion are translated as  
\begin{align}
\big\{\partial^2+0\cdot\frac{\kappa^2H^2}{4\pi^2}\log a(\tau)\partial^2\big\}\hat{\varphi}_w(x)\simeq 0, 
\label{WEs}\end{align}
\begin{align}
i\big\{\gamma^\mu\partial_\mu-\frac{1}{4}\frac{\kappa^2H^2}{4\pi^2}\log a(\tau)\gamma^i\partial_i\big\}
\hat{\psi}_w(x)\simeq 0.  
\label{WED}\end{align}
The result (\ref{WEs}) is consistent with the cancellation of the IR logarithms shown in \cite{Kahya2007}. 
Although (\ref{WED}) does not correspond with the result obtained in \cite{Miao2005(2),Miao2005(3)}, 
we should note that the authors of these papers focus on the dynamics at the super-horizon scale. 
By focusing on the dynamics at the sub-horizon scale, 
we can derive (\ref{WED}) from Eq. (229) in \cite{Miao2005(1)}.  

Eq. (\ref{WED}) breaks the effective Lorentz invariance. 
Since we have shown that the effective Lorentz invariance holds in our original quantization scheme, 
there should be a prescription to retain it in a different quantization scheme. 
There must be a reasoning to select which quantization scheme should be chosen for the description of physics. 
We address these questions in the next section. 

\section{Prescription to retain the effective Lorentz invariance}
\setcounter{equation}{0}

In investigating how to retain the effective Lorentz invariance, 
we deal with the different parametrization of the metric and the matter field redefinition separately. 
First, let us consider the violations of the effective Lorentz symmetry due to the different parametrization of the metric. 
We should remember that the parametrization dependence of the metric emerges 
only in the tadpole diagrams (\ref{Ps1}), (\ref{PD1}). 
So we can compensate them by introducing the classical expectation value of the background metric. 
Note that the gravitational action is stationary with this shift.  
\begin{align}
h_{\mu\nu}\to v_{\mu\nu}+h_{\mu\nu},\hspace{1em}v^\mu_{\ \mu}=0, 
\label{v}\end{align}
\begin{align}
\kappa v_{\mu\nu}=-\kappa \langle h_{\mu\nu}\rangle|_\text{NL}
\simeq \frac{\kappa^2H^2}{4\pi^2}\log a(\tau)
\big\{\frac{3}{4}\delta_\mu^{\ 0}\delta_\nu^{\ 0}+\frac{1}{4}(\eta_{\mu\nu}+\delta_\mu^{\ 0}\delta_\nu^{\ 0})\big\}. 
\label{v1}\end{align}
At least at the one-loop level, 
the compensation by shifting the background metric is available 
not only for the difference between (\ref{para1})-(\ref{para3}) and (\ref{Wpara1})-(\ref{Wpara2}),  
but also for an arbitrary difference of the parametrization of the metric.  
It is because the difference at the non-linear level emerges only in the tadpole diagrams at the one-loop order.  

Next, we consider the violations of the effective Lorentz symmetry due to the matter field redefinition. 
Of the diagrams which contribute to (\ref{Fs1}) and (\ref{FD1}),  
the following ones break the effective Lorentz symmetry: 
\begin{align}
&\partial_\mu\big\{-2\kappa^2\langle w_+(x)(h^{\mu\nu})_+(x)\rangle\partial_\nu\hat{\varphi}_w(x)\big\} \\
&+i\kappa^2\partial_\mu\int d^4x'\ c_{AB}\partial_\sigma'\big\{
\big(-2\langle w_+(x)(h^{\rho\sigma})_A(x')\rangle\eta^{\mu\nu}-2\langle (h^{\mu\nu})_+(x)w_A(x')\rangle\eta^{\rho\sigma}\big) \notag\\
&\hspace{10em}\times\langle\partial_\nu(\varphi_w)_+(x)\partial_\rho'(\varphi_w)_B(x')\rangle\big\}\hat{\varphi}_w(x') \notag\\
\to&\ \frac{\kappa^2H^2}{4\pi^2}\log a(\tau)\big\{-\frac{3}{4}\partial_0^2-\frac{1}{4}\partial_i^2\big\}\hat{\varphi}_w(x), \notag
\end{align} 
\begin{align}
&-i\frac{3}{2}\kappa^2\langle w_+(x)(h^\mu_{\ a})_+(x)\rangle\gamma^a\partial_\mu\hat{\psi}_w(x) \\
&+i\kappa^2\int d^4x'\ c_{AB}\partial_\nu'\big\{\big(-\frac{3}{2}\langle w_+(x)(h^\nu_{\ b})_A(x')\rangle\eta^\mu_{\ a}
-\frac{3}{2}\langle (h^\mu_{\ a})_+(x)w_A(x')\rangle\eta^\nu_{\ b}\big) \notag\\
&\hspace{9em}\times\gamma^a\langle\partial_\mu(\psi_w)_+(x)(\bar{\psi}_w)_B(x')\rangle\big\}\gamma^b\hat{\psi}_w(x') \notag\\
\to&\ i\frac{\kappa^2H^2}{4\pi^2}\log a(\tau)\big\{\frac{9}{16}\gamma^0\partial_0-\frac{3}{16}\gamma^i\partial_i\big\}\hat{\psi}_w(x). \notag
\end{align} 
See Appendix \ref{A:A} for more details. 
Unlike the violations due to the different parametrization of the metric, 
we cannot compensate them altogether by shifting the background metric. 

In order to clarify the field redefinition dependence, 
we introduce parameters $\alpha_s$, $\alpha_D$ as follows
\begin{align}
\varphi_{\alpha_s}=a e^{\frac{2-\alpha_s}{2}\kappa w}\varphi,\hspace{1em}
\psi_{\alpha_D}=a^\frac{3}{2} e^{\frac{3-\alpha_D}{2}\kappa w}\psi. 
\label{Aparaf1}\end{align}
The corresponding Lagrangians are
\begin{align}
\mathcal{L}_s=-\frac{1}{2}e^{\alpha_s\kappa w}\tilde{g}^{\mu\nu}\partial_\mu\varphi_{\alpha_s}\partial_\nu\varphi_{\alpha_s}
-\frac{1}{2}(a e^{\frac{2-\alpha_s}{2}\kappa w})^{-1}\partial_\mu\big\{e^{\alpha_s\kappa w}\tilde{g}^{\mu\nu}
\partial_\nu (a e^{\frac{2-\alpha_s}{2}\kappa w})\big\}\varphi^2_{\alpha_s}, 
\label{Aparaf2}\end{align}
\begin{align}
\mathcal{L}_D=i\bar{\psi}_{\alpha_D}\gamma^a e^{\alpha_D\kappa w}\tilde{e}^{\mu}_{\ a}
\nabla_\mu|_{e^{\frac{2}{3}\alpha_D\kappa w}\tilde{g}_{\rho\sigma}}\psi_{\alpha_D}. 
\label{Aparaf3}\end{align}
The matter field redefinition (\ref{paraf1}) corresponds with
\begin{align}
\alpha_s=\alpha_D=0, 
\label{paraf4}\end{align}
and (\ref{Wparaf1}) corresponds with
\begin{align}
\alpha_s=2,\hspace{1em}\alpha_D=3.  
\label{Wparaf4}\end{align}
By the matter field redefinition $\varphi_0\to\varphi_{\alpha_s}$, $\psi_0\to\psi_{\alpha_D}$, 
the effective Lorentz invariance is broken by the following contributions
\begin{align}
&\ \partial_\mu\big\{-\alpha_s\kappa^2\langle w_+(x)(h^{\mu\nu})_+(x)\rangle\partial_\nu\hat{\varphi}_{\alpha_s}(x)\big\}\\
&+i\kappa^2\partial_\mu\int d^4x'\ c_{AB}\partial_\sigma'\big\{\big(-\alpha_s\langle w_+(x)(h^{\rho\sigma})_A(x')\rangle\eta^{\mu\nu}
-\alpha_s\langle (h^{\mu\nu})_+(x)w_A(x')\rangle\eta^{\rho\sigma}\big) \notag\\
&\hspace{10em}\times\langle\partial_\nu(\varphi_{\alpha_s})_+(x)\partial_\rho'(\varphi_{\alpha_s})_B(x')\rangle\big\}
\hat{\varphi}_{\alpha_s}(x') \notag\\
\to&\ \frac{\kappa^2H^2}{4\pi^2}\log a(\tau)\times\alpha_s\big\{-\frac{3}{8}\partial_0^2-\frac{1}{8}\partial_i^2\big\}
\hat{\varphi}_{\alpha_s}(x), \notag
\end{align} 
\begin{align}
&-i\frac{1}{2}\alpha_D\kappa^2\langle w_+(x)(h^\mu_{\ a})_+(x)\rangle\gamma^a\partial_\mu\hat{\psi}_{\alpha_D}(x) \\
&+i\kappa^2\int d^4x'\ c_{AB}\partial_\nu'\big\{\big(-\frac{\alpha_D}{2}\langle w_+(x)(h^\nu_{\ b})_A(x')\rangle\eta^\mu_{\ a} 
-\frac{\alpha_D}{2}\langle (h^\mu_{\ a})_+(x)w_A(x')\rangle\eta^\nu_{\ b}\big) \notag\\
&\hspace{9em}\times\gamma^a\langle\partial_\mu(\psi_{\alpha_D})_+(x)(\bar{\psi}_{\alpha_D})_B(x')\rangle\big\}
\gamma^b\hat{\psi}_{\alpha_D}(x') \notag\\
\to&\ i\frac{\kappa^2H^2}{4\pi^2}\log a(\tau)\times\alpha_D\big\{\frac{3}{16}\gamma^0\partial_0-\frac{1}{16}\gamma^i\partial_i\big\}
\hat{\psi}_{\alpha_D}(x). \notag
\end{align} 
These violations can be compensated together by shifting of the background metric only when
\begin{align}
\alpha_s=\alpha_D=\alpha. 
\label{A}\end{align}
In this case, including (\ref{v1}), the classical expectation value of metric is identified as 
\begin{align}
\kappa v_{\mu\nu}
=&\ \frac{\kappa^2H^2}{4\pi^2}\log a(\tau)
\big\{\frac{3}{4}\delta_\mu^{\ 0}\delta_\nu^{\ 0}+\frac{1}{4}(\eta_{\mu\nu}+\delta_\mu^{\ 0}\delta_\nu^{\ 0})\big\} \label{vA}\\
&+\frac{\kappa^2H^2}{4\pi^2}\log a(\tau)\times\alpha
\big\{-\frac{3}{8}\delta_\mu^{\ 0}\delta_\nu^{\ 0}-\frac{1}{8}(\eta_{\mu\nu}+\delta_\mu^{\ 0}\delta_\nu^{\ 0})\big\}. \notag
\end{align} 

As seen in (\ref{paraf4}) and (\ref{Wparaf4}), 
the matter field redefinition (\ref{paraf1}) belongs to the effective Lorentz invariant cases (\ref{A}) while (\ref{Wparaf1}) does not. 
We claim that the matter field redefinition (\ref{paraf1}) is the natural choice for the description of physics.  
To show our reasoning, let us consider the following self-interaction with the dimensionless couplings: 
\begin{align}
\mathcal{L}_4=-\frac{\lambda_4}{4!}\sqrt{-g}\varphi^4,\hspace{1em}
\mathcal{L}_Y=-\lambda_Y\sqrt{-g}\varphi\bar{\psi}\psi. 
\end{align}
In any case other than (\ref{paraf1}), these dimensionless couplings are scale dependent even at the classical level
\begin{align}
\mathcal{L}_4=-e^{2\alpha_s\kappa w}\frac{\lambda_4}{4!}\varphi_{\alpha_s}^4,\hspace{1em}
\mathcal{L}_Y=-e^{(\frac{1}{2}\alpha_s+\alpha_D)\kappa w}\lambda_Y\varphi_{\alpha_s}\bar{\psi}_{\alpha_D}\psi_{\alpha_D}. 
\end{align}
Furthermore, we should note that the kinetic terms are canonically normalized in the prescription (\ref{paraf1}). 
It simultaneously indicates that unitarity is respected \cite{Unz1985}. 
Thus unitarity also singles out this scheme. 
In field theory, unitarity and symmetry are intimately related as we find such a correspondence here in quantum gravity.

A sensible setting is that the kinetic terms are canonically normalized. 
In this setting, dimensionless couplings are scale dependent only when we consider quantum corrections to them. 
The choice of the matter field redefinition (\ref{paraf1}) is a unique way to satisfy such a requirement.    
Once we adopt this scheme, the remaining parametrization dependence of the metric can be eliminated 
by shifting the background metric (\ref{v}). 
In this way, we can obtain the results which preserve the effective Lorentz invariance. 
We point out that they do not depend on the parametrization of the metric.  

\section{Conclusion}
\setcounter{equation}{0}

The gravitational field on the dS background contains the scale invariant spectrum which is dominant at the super-horizon scale. 
The existence of the scale invariant spectrum indicates the sensitivity for a size of the universe. 
That is, the exponential expansion leads to the dS symmetry breaking term in the corresponding propagator.  

In the previous works \cite{KitamotoSD,KitamotoG}, 
we have investigated soft gravitational effects on the local dynamics of matter fields at the one-loop level. 
There, we have adopted the gauge fixing term (\ref{GF}), 
the parametrization of the metric (\ref{para1})-(\ref{para3}) and the matter field redefinition (\ref{paraf1}). 
In these settings, the effective Lorentz invariance of scalar, Dirac and gauge field theories are preserved and 
the couplings of $\phi^4$, Yukawa and gauge interactions are dynamically screened by soft gravitons. 
The preservation of the effective Lorentz invariance is true even when we deform the gauge fixing term slightly. 
Furthermore the time evolutions of these couplings are physical as their relative scaling exponents are gauge invariant. 

We extend our investigation of the IR effects on the local dynamics of matter fields in this paper. 
Specifically, we have clarified how the IR effects are depend on the quantization scheme: parametrization of the metric and the matter field redefinition. 
An arbitrary choice of the parametrization of the metric and the matter field redefinition 
do not allow the effective Lorentz invariance of the local dynamics. 
As for the parametrization dependence of the metric, we have shown that the violation terms of the effective Lorentz symmetry can be eliminated by shifting the background metric. 

In contrast, we cannot compensate the field redefinition dependence by the shift of the background metric. 
We have found that the effective Lorentz invariance can be retained only when we adopt the matter field redefinitions (\ref{A}). 
Our choice in the previous works (\ref{paraf1}) belongs to this case. 
In particular, we claim that our choice is the most preferable to describe physics. 
It is because only when we adopt the matter field redefinition (\ref{paraf1}), 
unitarity is respected as the kinetic terms are canonically normalized 
and all dimensionless couplings of self-interactions become scale independent at the classical level. 
In canonical quantization with a Hamiltonian, unitarity is manifest. 
However it is not so in a Lagrangian formalism since we need to worry about non-trivial Jacobians. 
In many cases, symmetry picks up a unitary theory such as gauge symmetry for  a gauge theory and reparametrization invariance for a non-linear sigma model. 
In our case, conformal invariance plays such a role. 

By selecting this matter field redefinition and shifting the background metric, 
the IR effects on the kinetic terms are equal to our results in the previous works \cite{KitamotoSD,KitamotoG}. 
That is, the IR effects on the kinetic terms preserve the effective Lorentz invariance and can be absorbed by the wave function renormalization factors. 
We conclude that there is no physical effect from IR logarithms in free field theories at the sub-horizon scale. 
This conclusion implies that there is no IR logarithmic effect in generating curvature perturbation when we study CMB. 
The IR divergences due to soft gravitons can be eliminated by a time dependent wave function renormalization. 
It is because the logarithmic time dependence can be regarded as constant in time at sub-horizon scale.
When we study IR effects in CMB, we still need to understand whether there is any logarithmic IR effect to curvature perturbation at the super-horizon scale while it classically stays constant.

Also for the IR effects on the dimensionless couplings, we can obtain the same results in the previous works. 
It is because the shift of the background metric contributes only to the tadpole diagrams at the one-loop level 
and we have only to consider soft gravitons running between the vertices except for the wave function renormalization factors. 

We should emphasize that our investigations in this paper and the previous works are up to the one-loop level. 
It is a non-trivial question whether the effective Lorentz invariance of the local dynamics and 
the gauge invariance of the scaling exponents of the effective couplings hold at higher loop levels.   
In particular, the shift of the background metric contributes not only to the tadpole diagrams 
but other types of diagrams at higher loop levels. 
In addition, we have not investigated the gauge dependences of the IR effects 
against large deformations of the gauge fixing term (\ref{GF}). 
The investigations of the IR effects in these regions are open issues. 

\section*{Acknowledgment}
This work is supported in part by the Grant-in-Aid for Scientific Research
from the Ministry of Education, Science and Culture of Japan. 
We would like to thank organizers and participants of the workshop "Physics of de Sitter Spacetime". 

\appendix
\section{Local terms with IR logarithms from (\ref{Fs1}), (\ref{FD1})}\label{A:A}
\setcounter{equation}{0}

In this appendix, we list the local contribution which comes from each term of (\ref{Fs1}) and (\ref{FD1}). 
In a similar way to (\ref{sEoM31})-(\ref{DEoM3}), the following local terms are extracted up to $\mathcal{O}(\log a(\tau))$: 
\begin{align}
\partial_\mu\big\{2\kappa\langle w_+(x)\rangle|_\text{NL}\eta^{\mu\nu}\partial_\nu\hat{\varphi}_w(x)\big\}
\simeq\frac{\kappa^2H^2}{4\pi^2}\log a(\tau)\times-\frac{3}{4}\partial^2\hat{\varphi}_w(x), 
\label{t1}\end{align}
\begin{align}
\partial_\mu\big\{2\kappa^2\langle w_+(x)w_+(x)\rangle\eta^{\mu\nu}\partial_\nu\hat{\varphi}_w(x)\big\}
\simeq\frac{\kappa^2H^2}{4\pi^2}\log a(\tau)\times-\frac{3}{8}\partial^2\hat{\varphi}_w(x), 
\label{t2}\end{align}
\begin{align}
\partial_\mu\big\{-2\kappa^2\langle w_+(x)(h^{\mu\nu})_+(x)\rangle\partial_\nu\hat{\varphi}_w(x)\big\}
\simeq\frac{\kappa^2H^2}{4\pi^2}\log a(\tau)\big\{\frac{3}{4}\partial_0^2+\frac{1}{4}\partial_i^2\big\}\hat{\varphi}_w(x), 
\label{t3}\end{align}
\begin{align}
&\ i\kappa^2\partial_\mu\int d^4x'\ c_{AB}\partial_\sigma'\big\{4\langle w_+(x)w_A(x')\rangle\eta^{\mu\nu}\eta^{\rho\sigma} 
\langle\partial_\nu(\varphi_w)_+(x)\partial_\rho'(\varphi_w)_B(x')\rangle\big\}\hat{\varphi}_w(x') \label{t4}\\
\to&\ \frac{\kappa^2H^2}{4\pi^2}\log a(\tau)\times\frac{3}{4}\partial^2\hat{\varphi}_w(x), \notag
\end{align}
\begin{align}
&\ i\kappa^2\partial_\mu\int d^4x'\ c_{AB}\partial_\sigma'\big\{
\big(-2\langle w_+(x)(h^{\rho\sigma})_A(x')\rangle\eta^{\mu\nu}-2\langle (h^{\mu\nu})_+(x)w_A(x')\rangle\eta^{\rho\sigma}\big) \label{t5}\\
&\hspace{9em}\times\langle\partial_\nu(\varphi_w)_+(x)\partial_\rho'(\varphi_w)_B(x')\rangle\big\}\hat{\varphi}_w(x') \notag\\
\to&\ \frac{\kappa^2H^2}{4\pi^2}\log a(\tau)\big\{-\frac{3}{2}\partial_0^2-\frac{1}{2}\partial_i^2\big\}\hat{\varphi}_w(x), \notag
\end{align}
\begin{align}
&-i\kappa^2\int d^4x'\ c_{AB}
\langle \partial^2w_+(x){\partial'}^2w_A(x')\rangle\langle(\varphi_w)_+(x)(\varphi_w)_B(x')\rangle\hat{\varphi}_w(x') \label{t6}\\
\to&\ \frac{\kappa^2H^2}{4\pi^2}\log a(\tau)\times-\frac{1}{8}\partial^2\hat{\varphi}_w(x), \notag
\end{align}
\begin{align}
i\cdot 3\kappa\langle w_+(x)\rangle|_\text{NL}\eta^\mu_{\ a}\gamma^a\partial_\mu\hat{\psi}_w(x)
\simeq i\frac{\kappa^2H^2}{4\pi^2}\log a(\tau)\times-\frac{9}{8}\gamma^\mu\partial_\mu\hat{\psi}_w(x), 
\label{t7}\end{align}
\begin{align}
i\frac{9}{2}\kappa^2\langle w_+(x)w_+(x)\rangle\eta^\mu_{\ a}\gamma^a\partial_\mu\hat{\psi}_w(x)
\simeq i\frac{\kappa^2H^2}{4\pi^2}\log a(\tau)\times-\frac{27}{32}\gamma^\mu\partial_\mu\hat{\psi}_w(x), 
\label{t8}\end{align}
\begin{align}
-i\frac{3}{2}\kappa^2\langle w_+(x)(h^\mu_{\ a})_+(x)\rangle\gamma^a\partial_\mu\hat{\psi}_w(x)
\simeq i\frac{\kappa^2H^2}{4\pi^2}\log a(\tau)\big\{-\frac{9}{16}\gamma^0\partial_0+\frac{3}{16}\gamma^i\partial_i\big\}\hat{\psi}_w(x), 
\label{t9}\end{align}
\begin{align}
&\ i\kappa^2\int d^4x'\ c_{AB}\partial_\nu'\big\{9\langle w_+(x)w_A(x')\rangle\eta^\mu_{\ a}\eta^\nu_{\ b} 
\gamma^a\langle\partial_\mu(\psi_w)_+(x)(\bar{\psi}_w)_B(x')\rangle\big\}\gamma^b\hat{\psi}_w(x') \label{t10}\\
\to&\ i\frac{\kappa^2H^2}{4\pi^2}\log a(\tau)\times\frac{27}{16}\gamma^\mu\partial_\mu\hat{\psi}_w(x), \notag
\end{align}
\begin{align}
&\ i\kappa^2\int d^4x'\ c_{AB}\partial_\nu'\big\{\big(-\frac{3}{2}\langle w_+(x)(h^\nu_{\ b})_A(x')\rangle\eta^\mu_{\ a}
-\frac{3}{2}\langle (h^\mu_{\ a})_+(x)w_A(x')\rangle\eta^\nu_{\ b}\big) \label{t11}\\
&\hspace{9em}\times\gamma^a\langle\partial_\mu(\psi_w)_+(x)(\bar{\psi}_w)_B(x')\rangle\big\}\gamma^b\hat{\psi}_w(x') \notag\\
\to&\ i\frac{\kappa^2H^2}{4\pi^2}\log a(\tau)\big\{\frac{9}{8}\gamma^0\partial_0-\frac{3}{8}\gamma^i\partial_i\big\}\hat{\psi}_w(x). \notag
\end{align}
By summing up (\ref{t1})-(\ref{t11}), we can derive (\ref{Fs2}) and (\ref{FD2}).
Note that (\ref{t3}), (\ref{t5}), (\ref{t9}) and (\ref{t11}) do not preserve the effective Lorentz invariance. 


\end{document}